\documentclass[journal,twocolumn]{IEEEtran}
\usepackage{amsfonts}
\usepackage{times}
\usepackage{graphicx}
\usepackage{latexsym}
\usepackage{dsfont}
\usepackage{amssymb}
\usepackage{amsmath}
\usepackage{cite}
\usepackage{verbatim}
\usepackage{subfigure}

\newcommand{\figref}[1]{{Fig.}~\ref{#1}}

\def\bb0{{\mathbb{0}}}

\def\ba{{\mathbf{a}}}
\def\bb{{\mathbf{b}}}

\def\bff{{\mathbf{f}}}

\def\bm{{\mathbf{m}}}

\def\b0{{\mathbf{0}}}

\def\bA{{\mathbf{A}}}

\def\bI{{\mathbf{I}}}

\def\bR{{\mathbf{R}}}

\def\bbE{{\mathbb{E}}}

\def\cA{\mathcal{A}}

\def\cN{\mathcal{N}}

\def\sf0{{\mathsf{0}}}

\usepackage{epstopdf}
\usepackage{enumerate}
\usepackage{algorithmicx}
\usepackage{algorithm}
\usepackage{amsmath}
\usepackage[noend]{algpseudocode}
\usepackage{float}
\usepackage{hyperref}
\usepackage{color}
\usepackage{makeidx}
\usepackage{bbm}
\usepackage{graphicx}
\usepackage{lipsum}
\usepackage{subfigure}
\usepackage{tablefootnote}
\usepackage{balance}

\def\j{\mathrm{j}}

\begin{document}

\title{Millimeter Wave Base Stations with Cameras: Vision Aided Beam and Blockage Prediction}
\author{Muhammad Alrabeiah\textsuperscript{*}, Andrew Hredzak\textsuperscript{*}\thanks{*Authors contributed equally}, and Ahmed Alkhateeb\\ Arizona State University, Emails: \{malrabei, ahredzak, alkhateeb\}@asu.edu}
\maketitle

\begin{abstract}
This paper investigates a novel research direction that leverages vision to help overcome the critical wireless communication challenges. In particular, this paper considers millimeter wave (mmWave) communication systems, which are principal components of 5G and beyond. These systems face two important challenges: (i) the large training overhead associated with selecting the optimal beam and (ii) the  reliability challenge due to the high sensitivity to link blockages. Interestingly,  most of the devices that employ mmWave arrays will likely also use cameras, such as 5G phones, self-driving vehicles, and virtual/augmented reality headsets. Therefore, we investigate the potential gains of employing cameras at the mmWave base stations and leveraging their visual data to help overcome the beam selection and blockage prediction challenges. To do that, this paper exploits computer vision and deep learning tools to predict mmWave beams and blockages directly from the camera RGB images and the sub-6GHz channels.  The experimental results reveal interesting insights into the effectiveness of such solutions. For example, the deep learning model is capable of achieving over 90\% beam prediction accuracy, which only requires snapping a shot of the scene and zero overhead.
\end{abstract}

\section{Introduction} \label{intro}
Massive number of antennas and high frequencies are the two dominating characteristics of future wireless communication technologies \cite{mmWaveIsFuture}. They both support the increasingly-high data rates that future technologies, like virtual reality, augmented reality, and autonomous driving, are demanding \cite{6GAndBeyond}. The transition towards high-frequency bands and large antenna arrays is evident in the new 5G technology; it adopts dual-band operation by using sub-6GHz and millimeter-wave (mmWave) bands \cite{Parkvall2017} and supports massive MIMO communications.

The increasing bandwidth and number of antennas do not come cheap. They bring with them a lot of control overhead that  prevents them from realizing their full potential. As future wireless communications move to mmWave and higher frequencies, propagation characteristics severely change \cite{CovCapaMmWvae}; mmWaves are known for their weak penetration ability and significant power loss when they reflect off surfaces. This places strong emphasis on the need for antenna directivity, which commands large antenna arrays, and Line-of-Sight (LOS) connections. Hence, beam-forming and blockage-prediction become critical tasks for any mmWave system. Both tasks are associated with large control overhead from the perspective of classical signal processing \cite{DetChPred, DLHybPreCod, CoordBeamForm}, which poses a major challenge to the support of \textit{mobility} and \textit{relaibility} in these systems. That control overhead has ignited a lot of interest in \textit{intelligent} (data-driven) solutions powered by machine learning and, in particular, its deep leaning paradigm. Examples of such solutions could be found in \cite{Alkhateeb2018a,CoordBeamForm, LIS, DetChPred, Sub6PredMmWave}.

The majority of the work adopting deep learning focuses on wireless sensory data to drive the learning and deployment of intelligent solutions, which begs the question of whether other forms of sensory data could be utilized to deal with the control overhead problem or not. Solutions like those in \cite{Eckelmann2017,Nakashima2018} provide a \textit{partially} positive answer to that question, where depth sensors are exploited to help  wireless communication objectives. In this work, we present \textit{Vision-Aided Wireless Communications} (VAWC) as a new wholistic paradigm to tackle the overhead problem. It ultimately utilizes not only depth and wireless data, but also RGB images to enable mobility and reliability in mmWave wireless communications.

The main objective of this paper is to present the promise and potential VAWC has by addressing the beam and blockage prediction tasks using RGB, sub-6 GHz channels, and deep learning. When a pre-defined beam-forming codebook is available, learning beam prediction from images degenerates to an image classification task; depending on the user location in the scene, each image could be mapped to a class represented by a unique beam index from the codebook. On the other hand, detecting blockage in still images could be slightly trickier than beams as the instances of no user and blocked user are visually the same. Hence, images are paired with sub-6 GHz channels to identify blocked users. Each problem is studied  in a single-user wireless communication setting. 

The rest of this paper is organized as follows. Section \ref{sys_ch}  presents the system and channel models used to study the beam and blockage prediction problems. Section \ref{prob_form} presents the formulation of the two problems. Following that, a detailed discussion on the proposed vision-aided beam and blockage prediction solutions  takes place in Section \ref{prop_sol}. The evaluation of both solutions is presented in Section \ref{sim_res}, which  starts with introducing the communication scenarios, benchmark dataset, and experimental setting. Then, it  presents the evaluation results. Finally, Section \ref{concl}  concludes this paper with some remarks and possible future work. 

\textbf{Notation}: We use the following notation throughout this paper: $\bA$ is a matrix, $\ba$ is a vector, $a$ is a scalar, $\cA$ is a set of scalars, and $\boldsymbol{\mathcal{A}}$ is a set of vectors. $\|\ba \|_p$ is the p-norm of $\ba$. $|\bA|$ is the determinant of $\bA$, whereas $\bA^T$, $\bA^{-1}$ are its transpose and inverse. $\bI$ is the identity matrix.  $\cN(\bm,\bR)$ is a complex Gaussian random vector with mean $\bm$ and covariance $\bR$.  $\bbE\left[\cdot\right]$ is used to denote expectation.


\section{System and Channel Models}  \label{sys_ch}
The following two subsections present the system and channel models adopted in this work.
\subsection{System model}

\begin{figure}
	\centering
	\subfigure[ ]{\includegraphics[width=.8\linewidth]{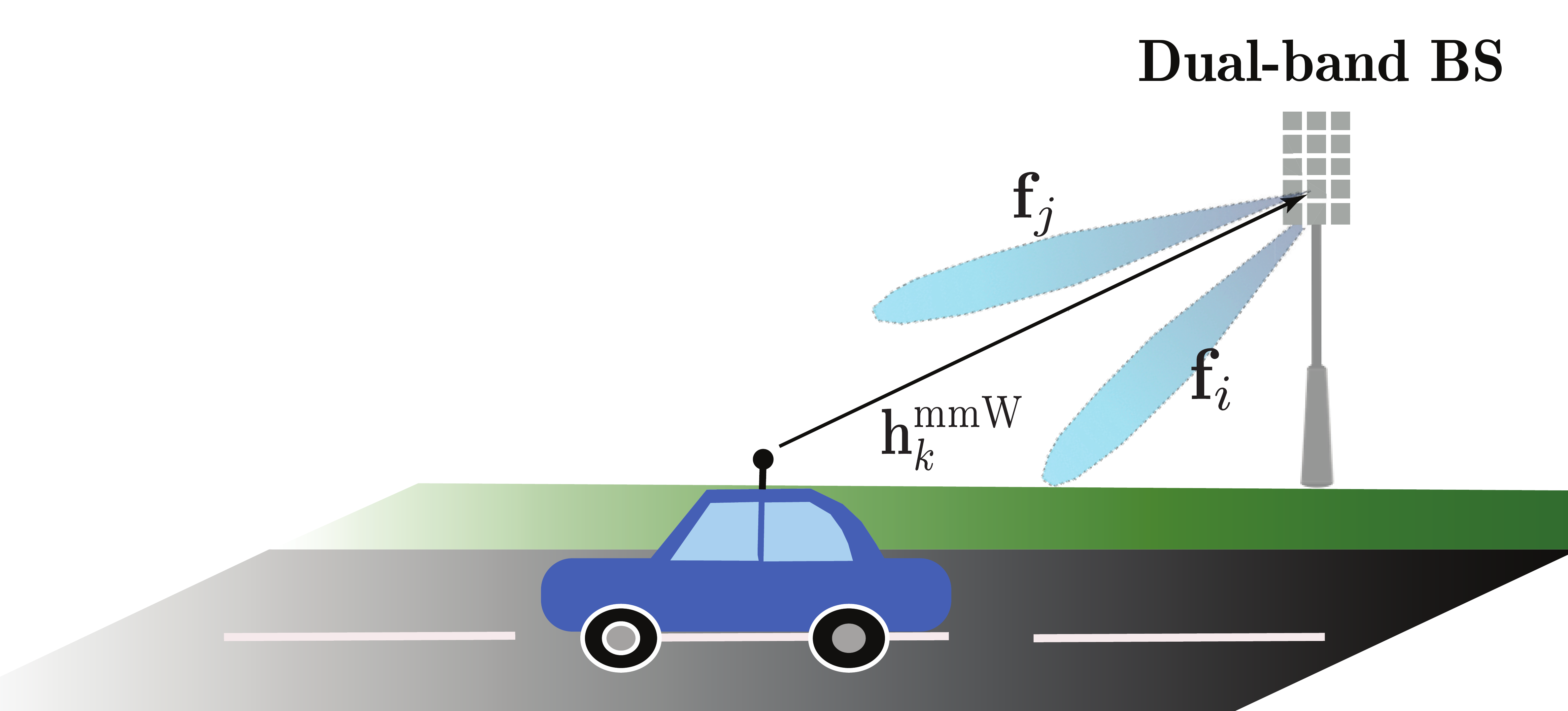}}
	\subfigure[ ]{\includegraphics[width=.8\linewidth]{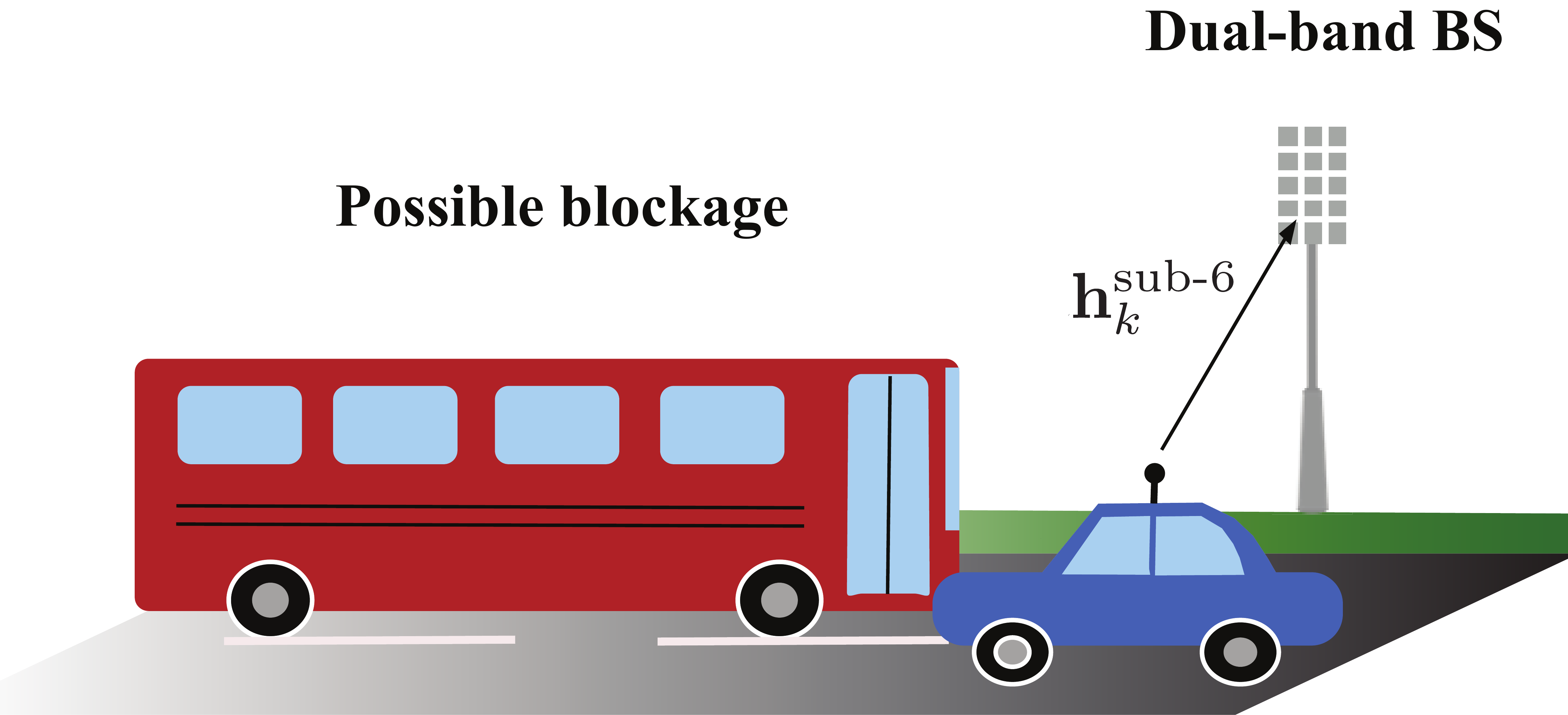}}
	\caption{Two system models are considered. (a) shows a downlink communication scenario where the Base Station (BS) is serving one user (the car) over the mmWave band. (b) shows an uplink communication scenario where control signaling  between the user and the BS happens over the sub-6 GHz band. The user, here, is getting closer to a possible blockage.}
	\label{sys_mod_fig}
\end{figure}

Consider a system where a Base Station (BS), operating at both sub-6GHz and mmWave bands, is communicating with a single-antenna  user, as depicted in \figref{sys_mod_fig}. The BS is assumed to be equipped with an $M_\text{mmW}$-element mmWave antenna array, an $M_\text{sub-6}$-element sub-6GHz antenna array, and an RGB camera. The system adopts  Orthogonal Frequency-Division Multiplexing (OFDM) with $K_\text{mmW}$ subcarriers at the mmWave band and a $K_\text{sub-6}$ subcarriers at sub-6GHz. Further, the mmWave BS systems is assumed to employ analog-only beamforming architecture while the sub-6GHz transceiver is assumed to be fully-digital \cite{HeathJr2016}. For mmWave beamforming, we assume that the beam is selected from a pre-defined beam codebook $\mathcal{F}=\{\mathbf{f}_1,\dots,\mathbf{f}_B\}$. To find the optimal beam, the user is assumed to send an uplink pilot that will be used to train the $B$ beams and select the one that maximizes the received power. This beam is then used for downlink data transmission. If beam $\bff_j$ is used in the downlink, then the received signal at the user's side can be expressed as
\begin{equation}\label{mmWSys}
  y_{k}^\text{mmW} = (\mathbf{h}^\text{mmW}_k)^T\mathbf{f}_js^\text{mmW}_k + n_k^\text{mmW},
\end{equation}
where $\mathbf{h}^\text{mmW}_k \in \mathbb{C}^{M_\text{mmW}\times1}$ 
is the mmWave channel of the $k$th subcarrier, $\mathbf{f}_j \in \mathbb{C}^{M_\text{mmW}\times1}$ is the $j$th beamforming vector in the codebook $\mathcal{F}$, $s_k^{\text{mmW}}$ is the symbol transmitted on the $k$th mmWave subcarrier, and $n_k^\text{mmW} \sim \mathcal N_{\mathbb C}(0,\sigma^2)$ is a complex Gaussian noise sample of the $k$th subcarrier frequency.

For blockage prediction, we assume that the BS will use the uplink signals on the sub-6GHz band for this objective.  If the mobile user sends an uplink pilot signal $s_k^\text{sub-6} \in \mathbb{C}$ on the $k$th subcarrier, then the received  signal at the BS can be written as
\begin{equation}\label{SubSys}
  \mathbf{y}^\text{sub-6}_k = \mathbf{h}^\text{sub-6}_k s^\text{sub-6}_k + \mathbf{n}^\text{sub-6}_k,
\end{equation}
where $\mathbf{h}^\text{sub-6}_k\in \mathbb{C}^{M_\text{sub-6}\times1}$ is the sub-6 GHz channel of the $k$th subcarrier, and $\mathbf{n}_k^\text{sub-6} \sim \mathcal{N}_{\mathbb C}(0,\sigma^2_\text{sub-6} \mathbf I)$ is the complex Gaussian noise vector of the $k$th subcarrier.

\subsection{Channel model}
This work adopts a geometric (physical) channel model  for the sub-6GHz and mmWave channels \cite{mmWaveIsFuture}. With this model, the mmWave channel (and similarly the sub-6GHz channel) can be written as: 
\begin{equation}
\mathbf{h}^{\text{mmW}}_k = \sum_{d=0}^{D-1} \sum_{\ell=1}^L \alpha_\ell e^{- \j \frac{2 \pi k}{K} d} p\left(dT_\mathrm{S} - \tau_\ell\right) \ba\left(\theta_\ell, \phi_\ell\right),
\end{equation} 
where $L$ is number of channel paths, $\alpha_\ell, \tau_\ell, \theta_\ell, \phi_\ell$ are the path gains (including the path-loss), the delay, the azimuth angle of arrival, and elevation, respectively, of the $\ell$th channel path. $T_\mathrm{S}$ represents the sampling time while $D$ denotes the cyclic prefix length (assuming that the maximum delay is less than $D T_\mathrm{S}$). Note that the advantage of the physical channel model is its ability to capture the physical characteristics  of the signal propagation including the dependence on the environment geometry, materials, frequency band, etc., which is crucial for considered beam and  blockage prediction problems.

\section{Problem Formulation} \label{prob_form}
Beam and blockage predictions are interleaved problems for any mmWave system. However, for the purpose of highlighting the potential of VAWC, they will be formulated and addressed separately in this work.

\begin{figure*}[t]
	\centering
	\includegraphics[width=\textwidth]{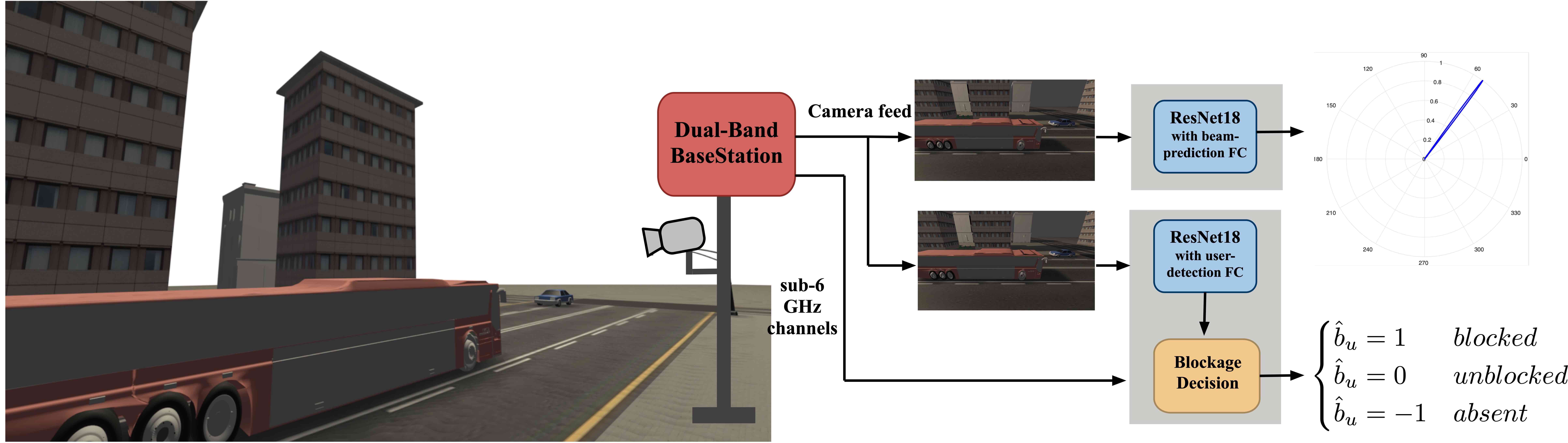}
	\caption{A block diagram of a vision-aided dual-band BS. Two ResNet18 models are deployed to learn beam prediction and user detection, respectively. Each network has a customized fully-connected layer that suits the task it handles. A network is trained to directly predict the beam index while the other predicts the user existence (detection) which is, then, converted to blockage prediction using the sub-6 GHz channels.}
	\label{sys_diag}
\end{figure*}

\subsection{Beam prediction}\label{beam_form}
The sole target of beam prediction is to determine the best beamforming vector $\mathbf f^*$ in the codebook $\mathcal F$ such that the received Signal-to-Noise-Ratio (SNR) at the receiver is maximized. In this work, the problem is viewed from a different perspective than that in the literature; the selection process depends on the camera feed instead of the explicit channel knowledge (i.e., $\mathbf{h}^{\text{mmW}}_k$) or beam training-- both requiring large overhead. Mathematically, this is expressed as follows:
\begin{equation}\label{opt_beam}
  \mathbf{f}^\star = \underset{\mathbf f \in \mathcal{F}}{\text{argmax}} \frac{1}{K}\sum_{k=1}^K \mathbb E \left[ \left\|(\mathbf{h}^{\text{mmW}}_k)^T \mathbf{f}\right\| _2^2\right],
\end{equation}
where $K$ is the total number of subcarriers. The optimal $\mathbf f^\star$, in this work, is found using an input image $X \in \mathbb R^{H\times W\times C}$, where $H$, $W$, and $C$ are, respectively, the hight, width, and number of color channels of the image. This is done using a \textit{prediction function} $f_\Theta(X)$ parameterized by a set of parameters $\Theta$ and outputs a probability distribution $\mathcal P = \{p_1, \dots, p_B \}$ over the vectors of $\mathcal F$. The index of the element with maximum probability determines the predicted beam vector, $\mathbf{\hat f} = \mathbf f_{j^\star}$, such that:
\begin{equation}
  j^\star = \underset{j\in\{1,\dots,B\}}{\text{argmax}}\left\lbrace p_1,\dots,p_j,\dots,p_B \right\rbrace.
\end{equation}
This function $f_\Theta(X)$ should be chosen to maximize the probability of correct prediction given an image $X$, i.e., $\mathbb P\left( \mathbf{\hat f} =\mathbf f^\star | X\right)$.

\subsection{Blockage prediction}\label{blk_form}
Determining whether a user's LOS link is blocked or not is a key task to boost reliability in mmWave systems. LOS status could be assessed based on some sensory data obtained from the communication environment. Examples of that are RGB images and sub-6 GHz channels, which are the sensory data of choice in this paper. Hence, let $(X,\mathbf h_u^\text{sub-6})$ be the pair of an RGB image of the scene and the user's sub-6 GHz channels, and let $b_u\in \{-1,0,1\}$ be the actual LOS status, where $1$, $0$, and $-1$ refer to the statuses: blocked link, unblocked link, and absent user. In similar spirit to beam prediction, the target of the system is to predict with high probability the status of the user $\hat{b}_u$ given $(X,\mathbf h_u^\text{sub-6})$ using a prediction function $G_\Theta(X,\mathbf h_u^{sub-6})$, which can be expressed with the following optimization problem:
\begin{equation}
  \underset{G_\Theta(X,\mathbf h_u^\text{sub-6})}{\text{max}} \prod_{u = 1}^U \mathbb P\left( \hat{b}_u = b_u | (X,\mathbf h_u^\text{sub-6})\right),
\end{equation}
where $U$ is the total number of user positions. Note here that the product of $\mathbb P\left( \hat{b}_u = b_u | (X,\mathbf h_u^\text{sub-6})\right)$ is a result of the assumption that the LOS status of a user position is conditionally independent from that of other positions. Despite that this assumption may not be accurate, it is a helpful simplification of the problem.

\section{Proposed Camera-Based Solutions} \label{prop_sol}
Two deep learning based solutions are proposed for the two problems. They both rely on deep convolutional networks and the concept of transfer learning. The cornerstone in each is the 18-layer Residual Network (ResNet-18) \cite{resnet} that is trained on the popular ImageNet2012 \cite{ImageNet12} and fine-tuned for the problem of interest. Figure \ref{sys_diag} depicts a block diagram of the two solutions, and the following two subsections present their details.

\begin{figure}[t]
	\centering
	\subfigure[ ]{\includegraphics[width=1\linewidth]{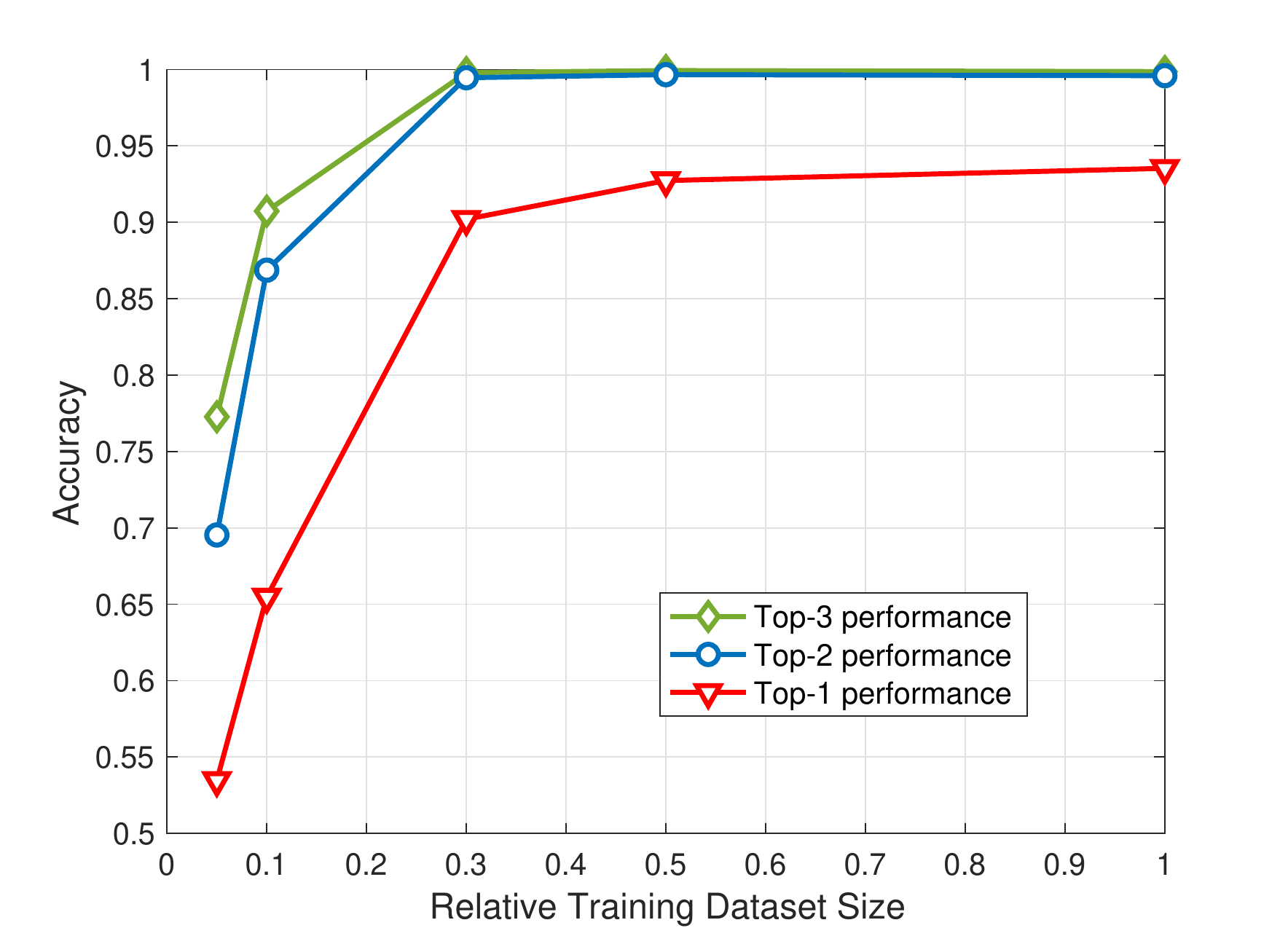}}
	\subfigure[ ]{\includegraphics[width=1\linewidth]{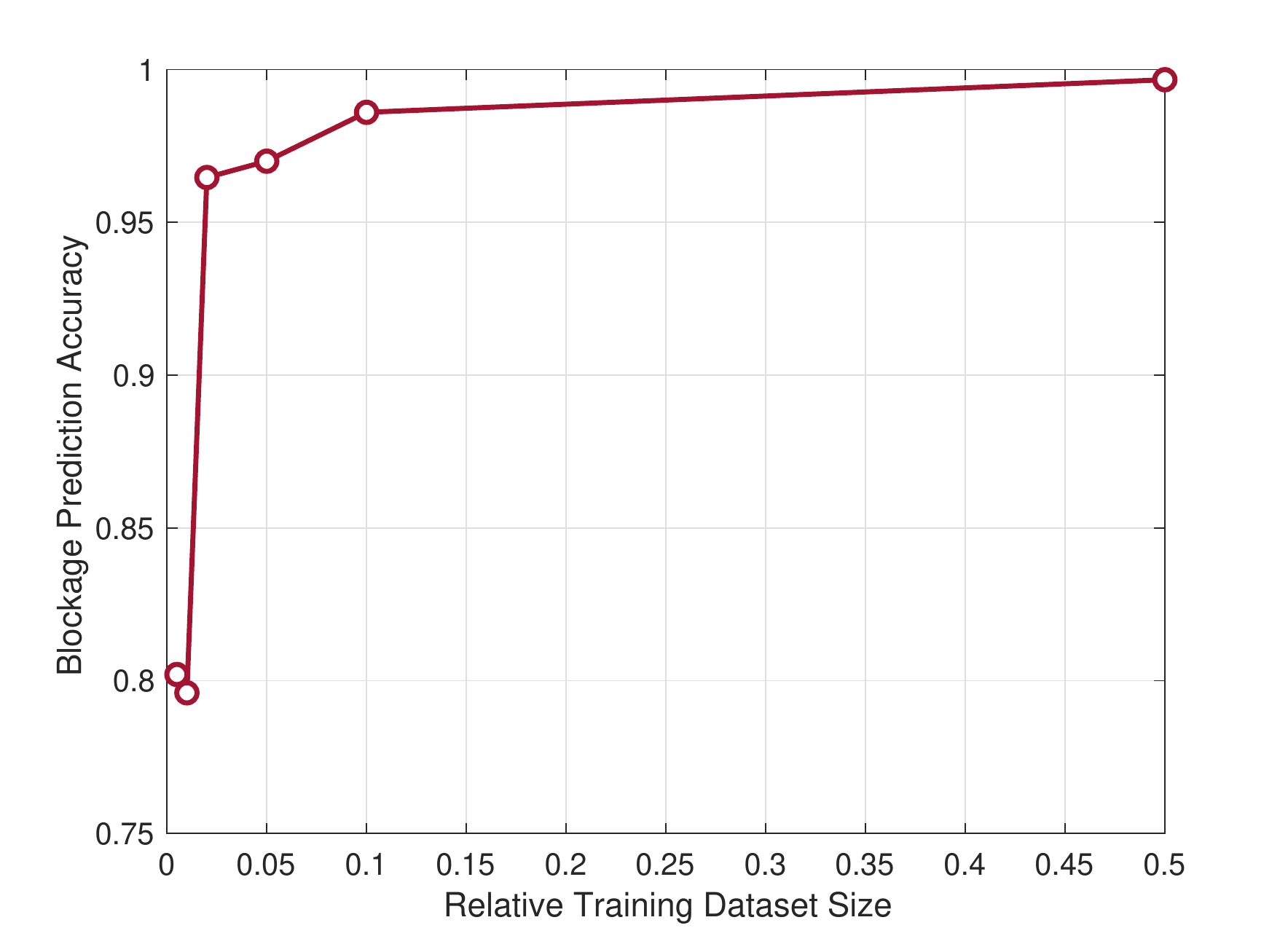}}
	\caption{The performances of the proposed solutions are shown in (a) and (b). The former shows the results for beam-prediction while the latter shows the results for user detection. Both figures present their respective accuracies versus relative training set size.}
	\label{perf_fig}
\end{figure}

\subsection{mmWave beam prediction}\label{beam_pred}
The idea of predicting the best beamforming vector from a codebook using an image has a strong analogy with image classification; the beam vectors divide the scene (spatial dimensions) into multiple sectors, and the goal of the system is to identify to which sector a user belongs. Clearly, assigning images to classes labeled by beam indices is possible in LOS situations as it relies on the knowledge of the user's location in the scene. Hence, the objective is to learn the class-prediction function $f_\Theta(X)$, see Section \ref{beam_form}, using images from the environment.

The proposed approach to learn the prediction function is based on deep convolutional neural networks and transfer learning. A pre-trained ResNet-18 model is adopted and customized to fit the beam prediction problem; its final fully-connected layer is removed and replaced with another fully-connected layer with a number of neurons equal to the codebook size, $B$ neurons. This model is then fine-tuned, in a supervised fashion, using images from the environment that are labeled with their corresponding beam indices. It basically learns the new classification function (i.e., $f_\Theta(X)$), that maps an image to a beam index. The training is conducted with a cross-entropy loss \cite{DLBook} given by:
\begin{equation}
  l = \sum_{i=1}^{B}t_i\log p_i,
\end{equation}
where $t_i$ is $1$ if $i$ is the beam index and $0$ otherwise. $p_i$ is the probability distribution induced by the soft-max layer.

\subsection{Link-blockage prediction}\label{blk_pred}
The blockage prediction problem is not very different from beam prediction in terms of the learning approach; it relies on detecting the user in the scene, and, thus, it could be viewed as a binary classification problem where a user is either detected or not. This, from a wireless communication perspective, is problematic as the absence of the user from the \textit{visual} scene does not necessarily mean it is blocked; it could simply mean that it does not exist. As a result, this paper proposes integrating images with sub-6 GHz channels to distinguish between absent and blocked users.

A valid question might arise at this point: why would the system not predict the link status from sub-6 GHz channels directly? This is certainly an interesting question, and the work in \cite{Sub6PredMmWave} has shown that neural networks can effectively learn blockage prediction from sub-6 GHz channels. However, a major issue with that approach is its need for labeled channels; there is no clear signal processing method for labelling sub-6 channels as blocked or not, and, on the other hand, labelling images is relatively easier. Therefore, a network trained to detect users could help predict blockages from still images when it is combined with sub-6 GHz channels. This approach could be used to label sub-6 GHz channels and use them later for training model like those in \cite{Sub6PredMmWave}. 

Blockage prediction here is performed in two stages: i) user detection using deep neural network, and ii) link status assessment using sub-6 GHz channels and the user-detection result. The neural network of choice for this task is also a ResNet-18 but with a 2-neuron fully-connected layer. Similar to Section \ref{beam_pred}, it is pre-trained on ImageNet data and fine-tuned on some images from the environment. It is first used to predict whether a user exists in the scene or not. If a user is detected, the link status is directly declared as unblocked. On the other hand, when the user is not detected, sub-6 GHz channels come into play to identify whether this is because it is blocked or it does not exist. When those channels are not zero, this means a user exists in the scene and it is blocked. Otherwise, a user is declared absent. 

\section{Simulation Results} \label{sim_res}
For the sake of emphasizing their potential, the two solutions are independently tested. Two datasets of synthetic data samples are used in these tests as, currently, there is no publicly-available dataset that combines real-world images and wireless channels. The following few subsections discuss the datasets, training of the neural networks, and their performance evaluation.

\begin{table}[t]
	\caption{Hyper-parameters for channel generation}
	\centering
	\begin{tabular}{|c | c | c|}
		\hline
		Parameter & \multicolumn{2}{c|}{value} \\
		\hline\hline
		Name of scenario & dist\textunderscore cam & colo\textunderscore cam\textunderscore blk \\
		\hline
		Active BSs & 3 & 1 \\
		\hline
		Active users & 1 to 5000 & 1 to 5000 \\
		\hline
		Number of antennas (x, y, x)  & (64,1,1) & (128,1,1) \\
		\hline
		System BW & 0.5 GHz & 0.5 GHz\\
		\hline
		Antenna spacing & 0.5 & 0.5 \\
		\hline
		Number of OFDM sub-carriers & 512 & 512\\
		\hline
		OFDM sampling factor & 1 & 1 \\
		\hline
		OFDM limit & 64 & 64\\
		\hline
	\end{tabular}
	\label{param}
\end{table}
\begin{table}[t]
	\caption{Hyper-parameters for network fine-tuning}
	\centering
	\begin{tabular}{|c | c | c|}
		\hline
		Parameter & \multicolumn{2}{c|}{value} \\
		\hline\hline
		Batch size & 150 & 150 \\
		\hline
		Learning rate & $1\times10^{-4}$ & $1\times10^{-4}$ \\
		\hline
		Weight decay & $1\times10^{-3}$ & $1\times10^{-3}$\\
		\hline
		Learning rate schedule & epochs 4 and 8 & epochs 4 and 8 \\
		\hline
		Learning-rate reduction factor & 0.1 & 0.1 \\
		\hline
		Data split (training-testing) & 70\%-30\% & 70\%-30\% \\
		\hline
	\end{tabular}
	\label{train_hyp}
\end{table}

\begin{figure*}[t]
	\centering
	\includegraphics[width=\linewidth]{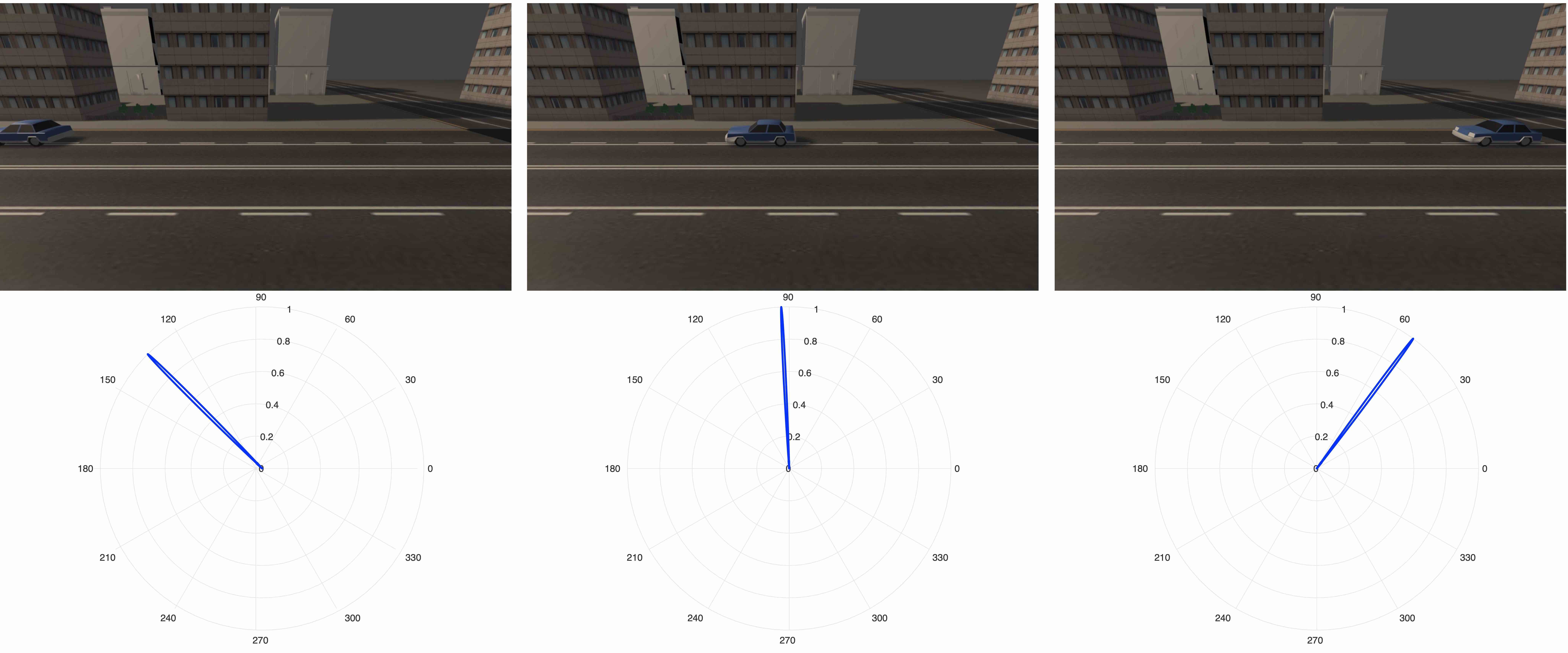}
	\caption{A visualization of the neural network inputs and outputs when it is deployed for beam prediction. For each column, the RGB image showing the location of the user (car in the image) is fed to the trained network, and the result is a beam index with the pattern shown below the image.}
	\label{car_beams}
\end{figure*}

\subsection{Scenario and datasets}\label{scena_data}
The publicly available ViWi framework \cite{Alrabeiah2019} is used to generate the datasets for testing the beam and blockage prediction solutions. ViWi provides four single-user communication scenarios and a data generator script. Two of those four are chosen for evaluation, namely the direct distributed-camera and blockage co-located-camera scenarios.

The direct distributed-camera scenario is used to generated data samples for the beam prediction experiments. The generated dataset has 5000 images and their corresponding mmWave channels; for each image depicting a user at some location, the corresponding mmWave channels of that user are generated using the generator package of ViWi. Table \ref{param} gives a summery of the channels generation hyper-parameters. An important point needs to be mentioned here. When generating the image-beam dataset, every image is paired with a beam from the codebook of the \textit{serving} BS, which is the one that \textit{sees} the user.

For blockage prediction experiments, the blockage co-located-camera scenario is used. A dataset of 5000 images is generated but without any mmWave or sub-6 GHz channels. The reason behind that lies in the role the neural network is playing in the blockage prediction solution. Its main job is to learn to recognize the user's existence, which only requires training with the RGB images of the scenario.

\subsection{Network training}\label{net_train}
For both experiments, ResNet-18 is customized by removing the last fully-connected layer and replacing it with either a 64-neuron (for beam prediction) or 2-neuron (for user detection) fully-connected layers. Each of the two new layers is initialized from a normal distribution with zero-mean and unit variance. The network, then, is fine-tuned on the training subset of one of the two datasets describe above. The training hyper-parameters, including the dataset split, are listed in Table \ref{train_hyp}. All training and testing was conducted on a system with a single NVIDIA RTX 2080Ti GPU using the popular PyTorch framework \cite{PyTorch}. Codes for the beam prediction experiment are made available at \cite{myGithub}.

\subsection{Prediction performance}\label{perf}
The ability of the neural network to predict beams from images is examined by studying the top-1, 2, and 3 accuracies\footnote{They are the complements of top-1, 2, and 3 errors commonly used as metrics for quantifying classification accuracy. See \cite{resnet} and \cite{ImageNet12}} versus the number of training samples. Figure \ref{perf_fig}-a shows the results of such test. The network shows good prediction performance with very little training samples, i.e., the accurate label is its first prediction around 90\% of the time after training with only 0.3 samples of the total training set size (1500 out of 3500). This gets improved further when the top-2 and 3 best predictions are considered; the accuracy jumps to almost 100\% with the same number of training samples. Top-1 accuracy continues to improve with more training data, and it hits 94\% when the whole training set is used. To give a better visualization of what these prediction accuracies mean, Figure \ref{car_beams} shows a sample of input images and their predicted beams.


For blockage prediction, the critical point is identifying the user's existence in the scene. As such, Figure \ref{perf_fig}-b depicts the user detection accuracy of a fine-tuned ResNet-18 versus training dataset size. It is evident that the network is capable of learning such task very well with little training; it requires a little less than 0.05 of the training samples (175 samples out of 3500) to produce an accuracy of around 96\%. Again, with more training samples, this accuracy approaches 100\%, e.g., in Figure \ref{perf_fig}-b, accuracy is around 99\% with half the training samples. 

 From a practical point of view, these numbers may not be very reflective if scenarios with varying (dynamic) environment and varying user shapes are considered. However, they hint at the great boost a mmWave system could get in supporting mobility and maintaining reliability shall visual-perception be incorporated, which is the objective of this paper (see Section \ref{intro}).
%

\section{Conclusion and Future Work} \label{concl}
Using computer vision and deep learning to tackle beam and blockage prediction problems is one promising approach to realize the potential of mmWave systems. The proposed solutions has clearly shown that promise for the case of single-user communications. Utilizing the strong correspondence between image classification and the tasks of beam prediction and user detection, a state-of-the-art deep learning model, like ResNet-18, trained for image classification could be fine-tuned to perform both tasks effectively. Both solutions need to be further developed and studied for dynamic environments with multiple users. These environments pose more difficult and realistic challenges to mmWave systems, and if the results of this paper are any indicator, vision-based approaches are definitely a strong contender for tackling those challenges.	 


\balance


\end{document}